\documentclass{goeproc}

\usepackage{graphics}
\usepackage{txfonts}
\usepackage{natbib}

\begin{document}

\title{Photospheric magnetic field and chromospheric emission}

\author{R.~Rezaei$^\ast$}
\author{R.~Schlichenmaier}
\author{C.~Beck}
\author{W.~Schmidt}

\affil{Kiepenheuer-Institut f\"ur Sonnenphysik, Freiburg, Germany}
\affil{$^{\ast}$\textit{Email:} rrezaei@kis.uni-freiburg.de}

\runningtitle{Photospheric magnetic field and chromospheric emission}
\runningauthor{R.~Rezaei et al.}

\firstpage{1}

\maketitle

\begin{abstract}
We present a statistical analysis of network and internetwork properties in the photosphere and the 
chromosphere. For the first time we simultaneously observed (a) the four Stokes parameters of the photospheric 
iron line pair at 630.2\,nm and (b) the intensity profile of the Ca H line at 396.8\,nm. The vector magnetic field 
was inferred from the inversion of the iron lines. We aim at an understanding of the coupling between 
photospheric magnetic field and chromospheric emission.
\end{abstract}

\section{Observations and data reduction}
We observed a series of 13 maps of a network region and the surrounding quiet Sun at a heliocentric angle of 
53$^\circ$, close to  the active region NOAA 10675 on September 27, 2004, 
with POLIS \citep{schmidt03,beck05b} at the German VTT in Tenerife.  
The Kiepenheuer Adaptive Optics System (KAOS) was used to improve spatial resolution to 
about 1 arcsec \citep{luhe_etal_03}.

Using \ the average profile \ of each map, we normalized \ the intensity at the line 
wing at \ \  396.490\,nm to the FTS profile \citep{stenflo_84}. 
>From the intensity profile of Ca H we define line properties like, e.g., the H-index, which is the integral 
around the line core from 396.8\,nm to 396.9\,nm.
The separation between network and internetwork is based on (i) the maps of magnetic flux density, (ii) the 
presence of Stokes-$V$ signals and emission in Ca H.

\begin{figure*}[t]
\center{\resizebox{14cm}{!}{\includegraphics*{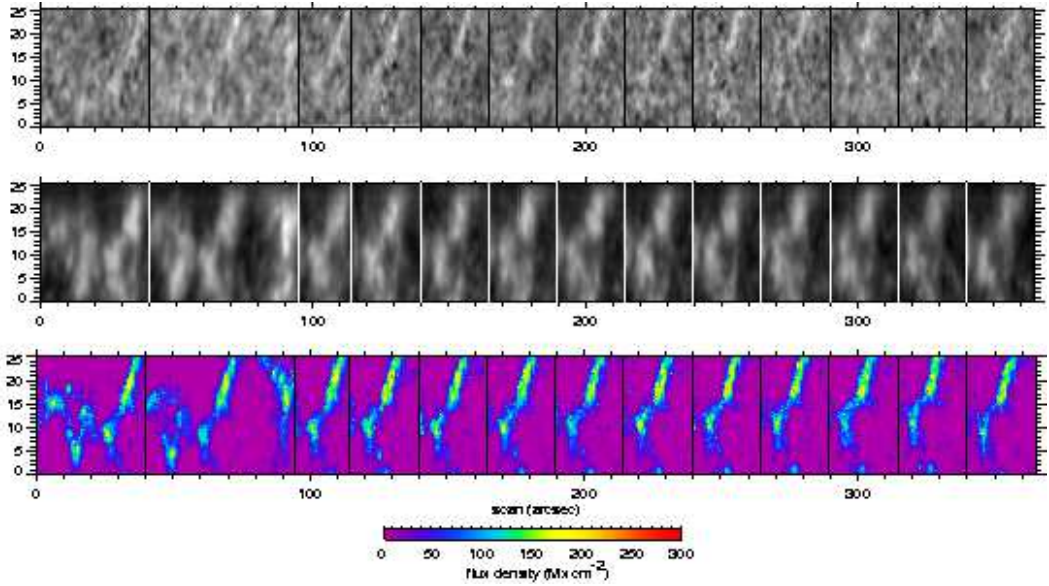}}}
\vspace{-5mm}
\caption{{\emph{ Top to buttom:}} the continuum intensity close to 630\,nm, the H-index, 
and the magnetic flux density obtained from the inversion.}
\end{figure*}

\section{Inversion}

An inversion was performed for the two iron lines at 630\,nm using the 
SIR code (Ruiz Cobo \& del Toro Iniesta 1992).  To mimic unresolved 
magnetic fields, we used a model atmosphere with one magnetic and one field-free component, plus stray light. 
The inversion yields a magnetic field vector, a line-of-sight velocity, and the magnetic flux per pixel. These 
quantities are constant along the line of sight. Using the flux density maps, we  created a  mask  to separate 
network  and  internetwork regions.

\begin{figure}[t]
\center\includegraphics[width=5.5cm]{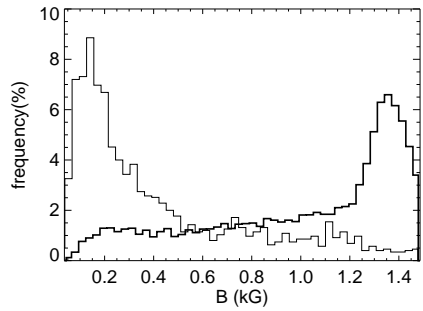}
\vspace{-1mm}
\caption{Distribution of the magnetic field strength for the network (thick) and internetwork (thin).}
\end{figure}

\section{Magnetic field distribution}
The polarization signal in $Q(\lambda)$, $U(\lambda)$, and $V(\lambda)$ is   
normalized by the local continuum intensity, $I_c$, for each pixel. 
The rms noise level of the Stokes parameters in the continuum  
was  $\sigma$\,=\,8.0\,$\times 10^{-4}$\,$I_c$ for the Fe\,I\,630\,nm lines.  
Only pixels with $V$ signals greater than 3\,$\sigma$  were 
included in the profile analysis. 
We obtain a magnetic field distribution which peaks at some 1.4\,kG for the network elements and at about 200\,G 
for the internetwork elements in agreement with previous infrared observations, but in contradiction with 
results from visible lines \citep{collados_01, lites_02}. 

\subsection{The effect of noise in the weak-field limit}
Our finding of weak rather than strong fields can be explained by the high spatial resolution and high 
polarimetric accuracy that we have achieved in our measurements. Since the internetwork magnetic fields are in 
the weak-field limit, the amount of noise strongly influences the outcome of the inversion. 
Figure 3 shows two different cases: 
\vspace{-1mm}
\begin{enumerate}
\item A fit of an original set of Stokes profiles yields $B$\,=\,840\,G  with a filling 
factor of 7.6\,\% (left panel, Fig. 3).
\item A fit on the same profile with added noise (such that the rms noise level is twice as large) 
delivers $B$\,=\,1.5\,kG and a filling factor of only 5.3\,\% (right four panels, Fig. 3).
\end{enumerate}

\begin{figure*}[t]
\center{\resizebox{12cm}{!}{\includegraphics{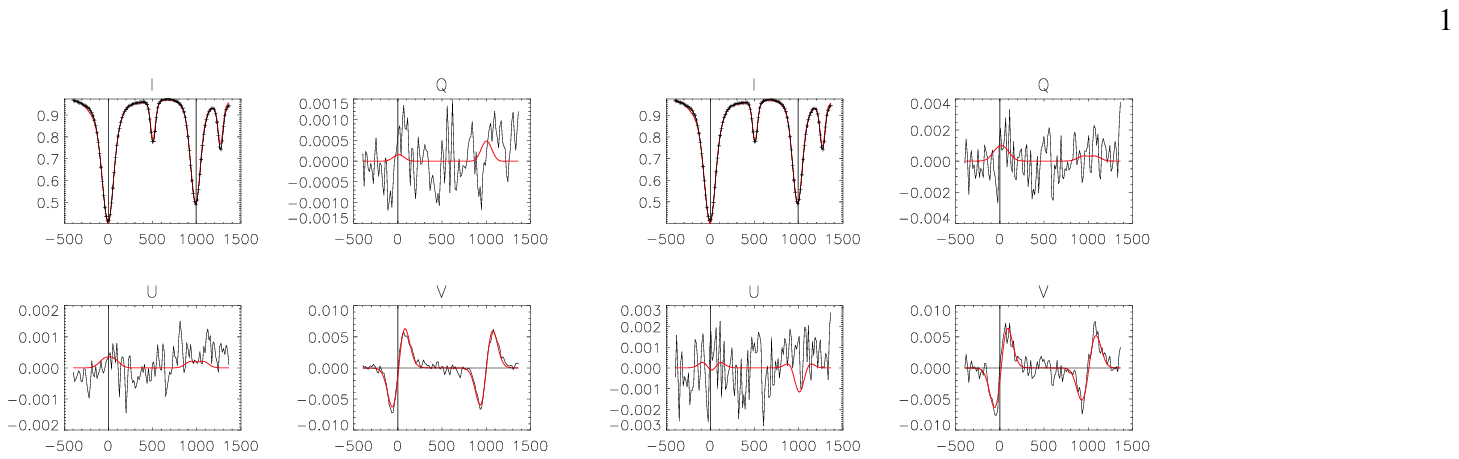}}}
\caption{The left (right) four panels show the inversion results of the original (noisy) data.}
\label{fig:calp1}
\end{figure*}

The reduction of noise and improvement of spatial 
resolution may resolve the existing discrepancy 
between visible and infrared magnetic field measurements.

\begin{figure*}[t]
\center{\resizebox{11cm}{!}{\includegraphics{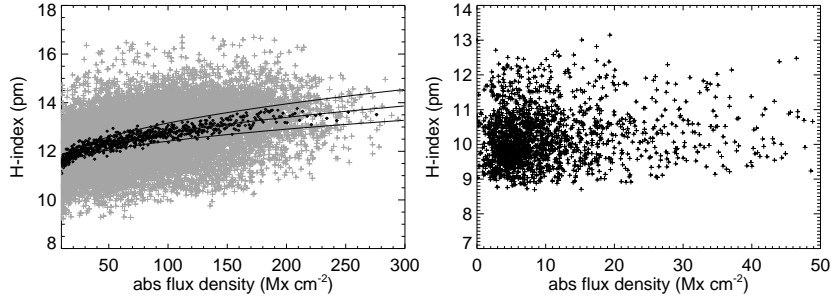}}}
\caption{The H-index vs. the magnetic flux density in the network (left) and internetwork (right). Black squares show 
the binned data. The curves show the fit to the original data along with 1\,$\sigma$ confidence level.}
\end{figure*}

\section{The H-index vs. magnetic flux density}

The left panel of Fig. 4 shows the H-index versus photospheric magnetic flux, $\Phi$, 
for the network. A power law fit, 
$H = a\,·\,\Phi^b + c$, yields  $b$\,=\,0.3 and $c$\,=\,10\,pm. 
In the internetwork the H-index does not correlate with the magnetic flux density 
(Fig. 4, right panel). 
The average value of the H-index in the internetwork is about $H$\,=\,10\,pm. 
The offset value, $c$\,=\,10\,pm, can be interpreted 
as the non-magnetic heating contribution and the stray light.

For the first time the H-index and simulaneously measured V-profile parameters can be compared. We find no 
correlation between these parameters in the internetwork (Fig. 5, pluses). In the network, a correlation 
exists, and the  H-index peaks  at  a small positive value for the area asymmetry and at vanishing V-profile 
Doppler shift (Fig. 5, squares).

\section{Conclusions}
$\bullet$ The internetwork magnetic field distribution peaks around 200\,G and  its mean absolute flux density 
amounts to 9\,Mx\,cm$^{-2}$. The finding of weak rather than strong fields is a consequence of the high spatial 
resolution and high polarimetric accuracy. 

\noindent $\bullet$ The H-index in the network is correlated to the magnetic flux density, approaching a value 
of $H$\,=\,10\,pm for vanishing flux.

\noindent $\bullet$ The H-index in the internetwork is not correlated to any property of the photospheric 
magnetic field implying that the chromospheric brightenings in the internetwork are non-magnetic.

\noindent $\bullet$ For high values of the H-index, the network shows small positive area (and amplitude) 
asymmetry, being consistent with the scenario of a line-of-sight crossing the magnetic boundary (canopy) of 
flux tubes that fans out with height \citep{steiner99}.

\begin{figure*}[t]
\center{\resizebox{11cm}{!}{\includegraphics{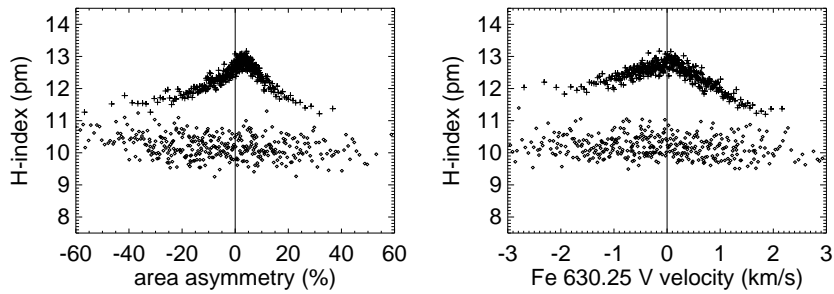}}}
\caption{Scatter plot of the area asymmetry and Stokes-$V$ velocity against the H-index.}
\label{fig:calp2}
\end{figure*}
\newcommand{\apj}{ApJ}%
\newcommand{\aap}{A\&A}%
\newcommand{\apjl}{ApJ}
\newcommand{\astnach}{Astr. Nachrichten}

\bibliographystyle{aa}
\bibliography{rezaei.bbl}

\end{document}